\documentclass[a4paper,11pt]{article}
\usepackage{pos,booktabs}

\title{MeV astronomy with \textsc{Herwig}?}

\author*[a,b]{Peter Reimitz}

\affiliation[a]{Institut f{\"u}r Theoretische Physik,
 Heidelberg, Germany}
 
\affiliation[b]{Instituto de Física, Universidade de São Paulo,
São Paulo, Brazil}

\emailAdd{peter@if.usp.br}

\abstract{GeV-scale dark matter is an increasingly attractive target for direct detection, indirect detection, and collider searches. Especially for masses in the MeV to GeV range, indirect detection is expected to give a leading constraint. In that range, dark matter annihilations into hadronic final states produce a challenging zoo of light hadronic resonances. With an update of \textsc{Herwig7}, we provide a modeling of these processes and study energy spectra from annihilation through a vector mediator. We cover dark matter masses between 250 MeV and 5 GeV and include an error estimate. This opens up new opportunities for fully studying indirect detection signals for hidden dark matter sectors with vector mediators in the sub-GeV range.}

\FullConference{%
  Tools for High Energy Physics and Cosmology - TOOLS2020\\
  2-6 November, 2020\\
  Institut de Physique des 2 Infinis (IP2I), Lyon, France}

\begin{document}
\maketitle

\section{Introduction}

Astrophysical and cosmological observations have shown that the bulk of the energy budget in the Universe is made out of dark matter (DM) and dark energy~\cite{Ade:2015xua}. Driven by the tremendous success of the Standard Model (SM) of particle physics and standard cosmology after Big Bang, it seems inevitable to look for a fundamental description of dark matter in terms of a quantum field theory. Any SM extension via operators coupling SM degrees of freedom to dark sector fields brings along a variety of ways to search for DM~\cite{Lin:2019uvt}. 

While indirect and direct detection as well as collider searches have set stringent constraints on weak-scale DM~\cite{Arcadi:2017kky}, processes in the sub-GeV range are comparably unexplored by standard searches. Alongside strong efforts in direct detection searches~\cite{Agnes:2018ves,Aprile:2016wwo,Essig:2017kqs}, one of the leading constraints in the MeV to GeV range is expected to be set by indirect detection~\cite{Leane:2018kjk}. Besides the existing GeV-scale gamma-ray and cosmic-ray observatories~\cite{Atwood:2009ez,Aleksic:2014lkm,Abramowski:2014tra,Holder:2006gi,Aguilar:2016vqr}, several proposed MeV gamma-ray telescopes such as e-\textsc{Astrogam}~\cite{DeAngelis:2017gra} and \textsc{AMEGO}~\cite{McEnery:2019tcm} and concept telescopes~\cite{Moiseev:2015lva,Duncan:2016zbd,Wu:2015wol,Hunter:2010} herald a new era in astronomy. A wealth of hidden sector models in the sub-GeV range can be tested. With additional degrees of freedoms in the dark sector compared to the standard WIMP, MeV DM annihilations and indirect detection signals often deviate from the weak-scale WIMP case. For example, a sizable fraction of DM annihilations can go to invisible particles~\cite{Leane:2018kjk}, DM can be resonantly produced~\cite{Bernreuther:2020koj}, the processes accounting for the DM abundance and the indirect detection signal are not described by one interaction only~\cite{DEramo:2018khz}, DM can be slightly but not fully asymmetric~\cite{Lin:2011gj}, or the emission of dark photons in DM bound state formations or transitions can cause an indirect detection signal~\cite{Baldes:2020hwx}.

Hence, both from a theoretical and experimental standpoint, a detailed theoretical study of DM annihilation processes in the sub-GeV range is of great importance. In particular, the calculation of the particle spectrum of DM annihilations is needed as a direct input to set indirect detection constraints on sub-GeV DM models.
Whereas for GeV to TeV DM we can rely on state-of-the-art codes, such as \textsc{PPPC4DMID}~\cite{Cirelli:2010xx}, \textsc{micrOMEGAs}~\cite{Belanger:2001fz,Belanger:2018ccd}, \textsc{MadDM}~\cite{Backovic:2013dpa,Arina:2020kko}, and \textsc{DarkSUSY}~\cite{Gondolo:2004sc,Bringmann:2018lay} based on simulations of high-energy collisions of elementary SM particles by \textsc{Pythia}~\cite{Sjostrand:2007gs,Sjostrand:2014zea} and ~\textsc{Herwig}~\cite{Bellm:2015jjp,Bellm:2019zci}, an implementation of sub-GeV annihilations including the decay of unstable final states is not available, neither for $e^+e^-$ collisions nor for DM processes. Processes at that energy scale can no longer be described by a perturbative hard process, followed by parton showers, fragmentation and the decay of hadronic states into stable final state particles. In the sub-GeV range the degrees of freedom are no longer quarks but hadrons. In the so-called vector meson dominance (VMD) model~\cite{OConnell:1995nse}, vector mesons act as mediators coupling to pions and kaons, and isospin serves as a guiding symmetry to describe processes in that energy range.

In the following\footnote{The majority of the material presented here is based on Ref.~\cite{Plehn:2019jeo}, which we refer the reader to for detailed discussions and an expanded list of references.}, we introduce the photon and positron spectra from (sub-) GeV DM annihilating into hadronic final states. First, we will explain how to calculate processes in the sub-GeV region once we are dealing with mesons instead of quarks. We will then use $e^+e^-$ data to fix the parametrization of the hadronic currents. The fit uncertainties can then be used to estimate the errors on the spectra. The processes are implemented into \textsc{Herwig}. With the current \textsc{Herwig}-based implementation, one can fully study DM annihilations and indirect detection constraints for vector mediator models. We comment on how our results can be provided for DM tools, such as \textsc{DarkSUSY} or \textsc{Hazma}~\cite{Coogan:2019qpu}.

\section{Processes in the sub-GeV range}
\begin{figure}[t]
\begin{center}
\includegraphics[width=.6\textwidth]{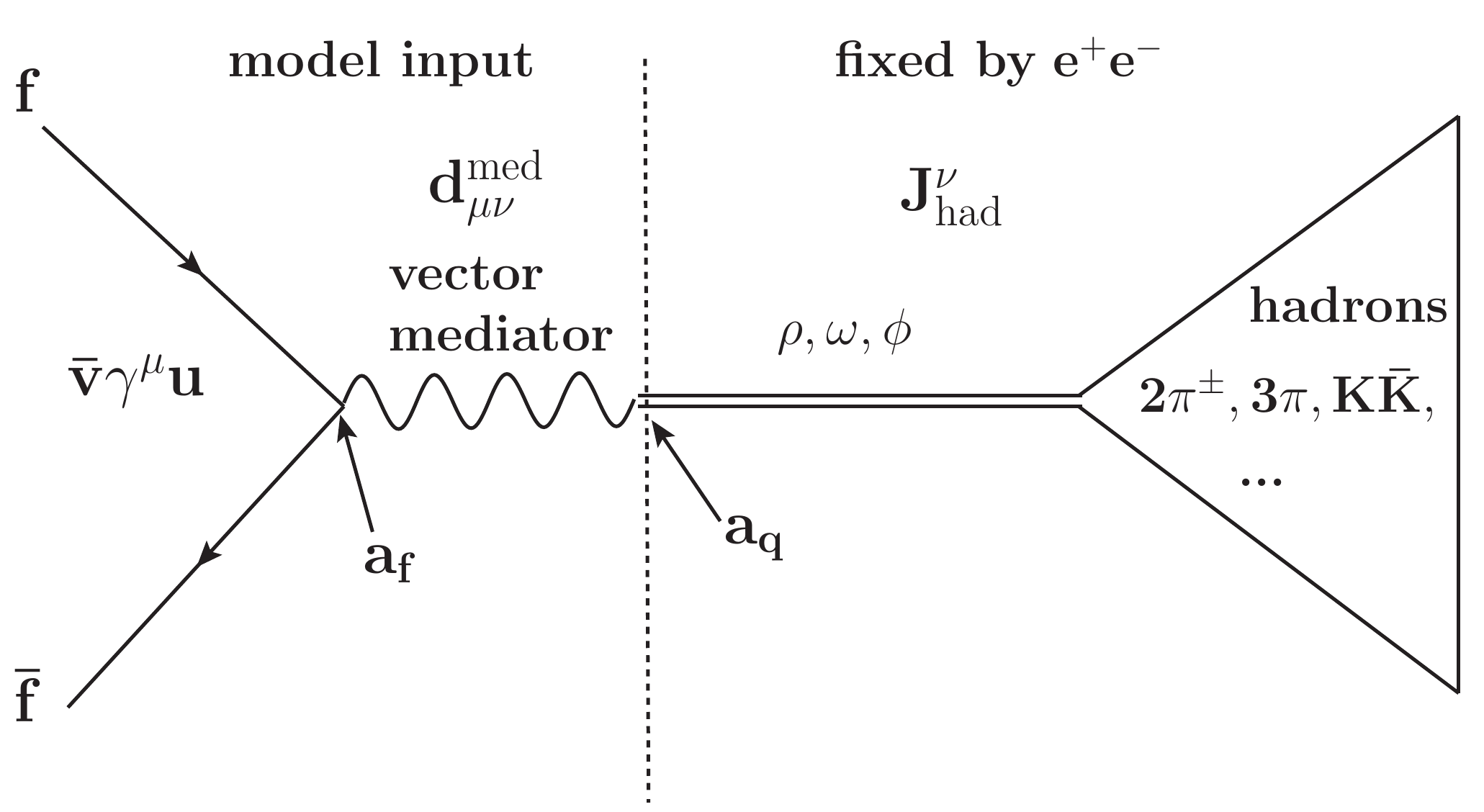}
\end{center}
\caption{\label{fig:LEprocess} Diagramatical structure of vector-mediated annihilation processes in the sub-GeV range}
\end{figure}

State-of-the-art Monte-Carlo (MC) collider tools such as \textsc{Pythia} and \textsc{Herwig} can describe DM annihilations into all sorts of final states for energy scales of several TeV down to 10 GeV~\cite{Cirelli:2010xx,Cembranos:2013cfa,Niblaeus:2019ldk,Amoroso:2018qga,Amoroso:2020mjm}. These MC-simulations split the annihilation process into a perturbative calculation of a matrix element, followed by a parton shower, both evaluated at energies around the center-of-mass energy of the process, down to a few GeV. For lower energies, we enter a non-perturbative regime where particles hadronize and form hadrons, that subsequently decay into other particles. Unlike in collider searches where neutrons, charged pions, muons and long-lived kaons are stable on collider scales, the only stable particles on astronomical scales are electrons, protons, photons, and neutrinos. In indirect detection, one important particle physics input is the energy spectrum of those stable final particles. As described in Ref.~\cite{Plehn:2019jeo}, the energy spectra produced by all standard DM tools agree with each other down to $\sim 5$~GeV in DM mass, corresponding to center-of-mass energies of 10 GeV. Below, we cannot trust these results anymore. The reason is that processes in the sub-GeV range are plagued by hadronic resonances. From the theoretical standpoint, the calculations of annihilations can no longer be composed of perturbative and non-perturbative calculations. Since the center-of-mass energy of the process is comparable to a scale where hadrons form, the degrees of freedom in our process are no longer quarks, but mesons and baryons. This hadronic structure is encoded in the hadronic current~\cite{Plehn:2019jeo} 
\begin{align}
J^\nu_{\rm had}=\sum_{q=u,d,s}a_q \bar{q}\gamma^\nu q=\frac{1}{\sqrt{2}}\left(
(a_u-a_d)\underbrace{J^{I=1,3,\nu}}_{\rho}+(a_u+a_d)\underbrace{J^{I=0,\nu}}_{\omega} \right)+a_s \underbrace{J^{s,\nu}}_{\phi}~.
\label{eq:current}
\end{align} 
It can be decomposed into an isospin $I=0$ and $I=1$ current, as well as a strange-quark current
\begin{align}
J^{I=1,3}_\nu&=\frac{1}{\sqrt{2}}(\bar{u}\gamma_\nu
               u-\bar{d}\gamma_\nu d),\nonumber\\
J^{I=0}_\nu&=\frac{1}{\sqrt{2}}(\bar{u}\gamma_\nu u+\bar{d}\gamma_\nu
             d),\nonumber\\
J^{s}_\nu &= \bar{s}\gamma_\nu s~.\nonumber
\end{align}
They represent the $\rho$ meson, $\omega$ meson and $\phi$ meson, respectively, and include excited states of those mesons as well. The intermediate mesons are related to the final state hadrons via isospin relations. Pure isospin $I=0$ channels are, for example, 
\begin{align}
\chi \chi \to \omega\pi\pi, \eta\omega, \ldots \, .
\label{eq:I0}
\end{align}
and the $I=1$ channel is producing
\begin{align}
\chi \chi \to \pi\pi, 4\pi, \eta\pi\pi, \omega\pi, \phi\pi, \eta'\pi\pi, \ldots
\label{eq:I1}
\end{align}
through the $\rho$ meson.
A sketch of the process structure can be seen in Fig.~\ref{fig:LEprocess}.
Depending on the quark couplings $a_q$ of the vector mediator that relates the initial current to the intermediate vector meson, one or the other part of the hadronic current might be dominant, sub-dominant, or even absent. The matrix element for these processes can be written as
\begin{align}
    \mathcal{M} = a_f\bar{v}(p_{\bar{f}})\gamma^\mu
u(p_{f}) d_{\mu\nu}^{\mathrm{med}}\langle X|J_{\text{had}}^\nu|0\rangle
\label{eq:ME}
\end{align}
for a fermionic incoming particle with vector mediator couplings $a_f$ and a vector mediator propagator $d_{\mu\nu}^{\rm med}$ and hadronic final states $X=2\pi,3\pi, K\bar{K},...$. From Fig.~\ref{fig:LEprocess} and Eq.~\eqref{eq:ME} we see that the process can be split into a DM model input and a part that does not involve any dark degrees of freedom. The model-dependent DM part consists of an incoming fermionic current and a vector mediator with couplings to quarks. Contrary to that, the isospin and strange-quark pieces of the hadronic current are independent of the DM model. They are based on low-energy QCD parametrizations in the VMD model and can be determined from $e^+e^-$ measurements. Hence, we can use the modeling of those DM model-independent currents as used in the description of $e^+e^-$ annihilations and transfer them to DM processes. A plethora of measurements of $e^+e^-$ annihilations into hadronic final states  in the sub-GeV range serve as a data input to fits to hadronic current parametrizations. We make use of existing modeling of annihilation channels into hadronic final states including two pseudoscalar mesons, a pseudoscalar meson plus a photon, and pseudoscalars plus vector mesons. Many channels are based on the low-energy tool \textsc{Phokhara}~\cite{Rodrigo:2001kf,Czyz:2017veo} and SND fits~\cite{Achasov:2006dv,Achasov:2016zvn}. For channels that have not been described yet, we introduce new parametrizations following the same theoretical background of VMD modelling. A complete list of all channels and its parametrizations and fits can be found in Tab.~\ref{tab:channels}. References for all parametrizations and fits can be found in Ref.~\cite{Plehn:2019jeo}.

\begin{table}[t!]
\centering
\begin{tabular}{l|ccc}
\hline
Channel & Parametrization & Fit & threshold energy [GeV]\\ \toprule
$\pi\gamma$ & SND& SND &0.140  \\ \midrule
$\pi\pi$ & Phokhara & Phokhara & 0.280\\
$\pi\pi\pi$ & Phokhara& Phokhara & 0.420\\
$\eta\gamma$ & SND & SND & 0.548\\
$4\pi$'s & Phokhara& own &0.560\\
$\eta\pi\pi$ &  Phokhara & own & 0.827\\
$\omega\pi$ & SND & SND &0.918\\
$KK$ & Phokhara & own & 0.996\\
$\omega\pi\pi$ &own& own & 1.062\\
$KK\pi$ &own&own & 1.135\\
$\phi\pi$ & own & own & 1.160\\
$\eta' \pi\pi$ & Phokhara & own & 1.237\\
$\eta\omega$& own & own & 1.331\\
$\eta\phi$  & own & own & 1.568\\
$p\bar{p}/n\bar{n}$ & Phokhara & own & 1.877\\
\bottomrule
\end{tabular}
\caption{Dominant processes contributing to $e^+e^- \to X$
 production in the relevant energy range. }
\label{tab:channels}
\end{table}

\section{Herwig4DM spectra}
\begin{figure}[th]
\centering
\includegraphics[width=0.49\textwidth]{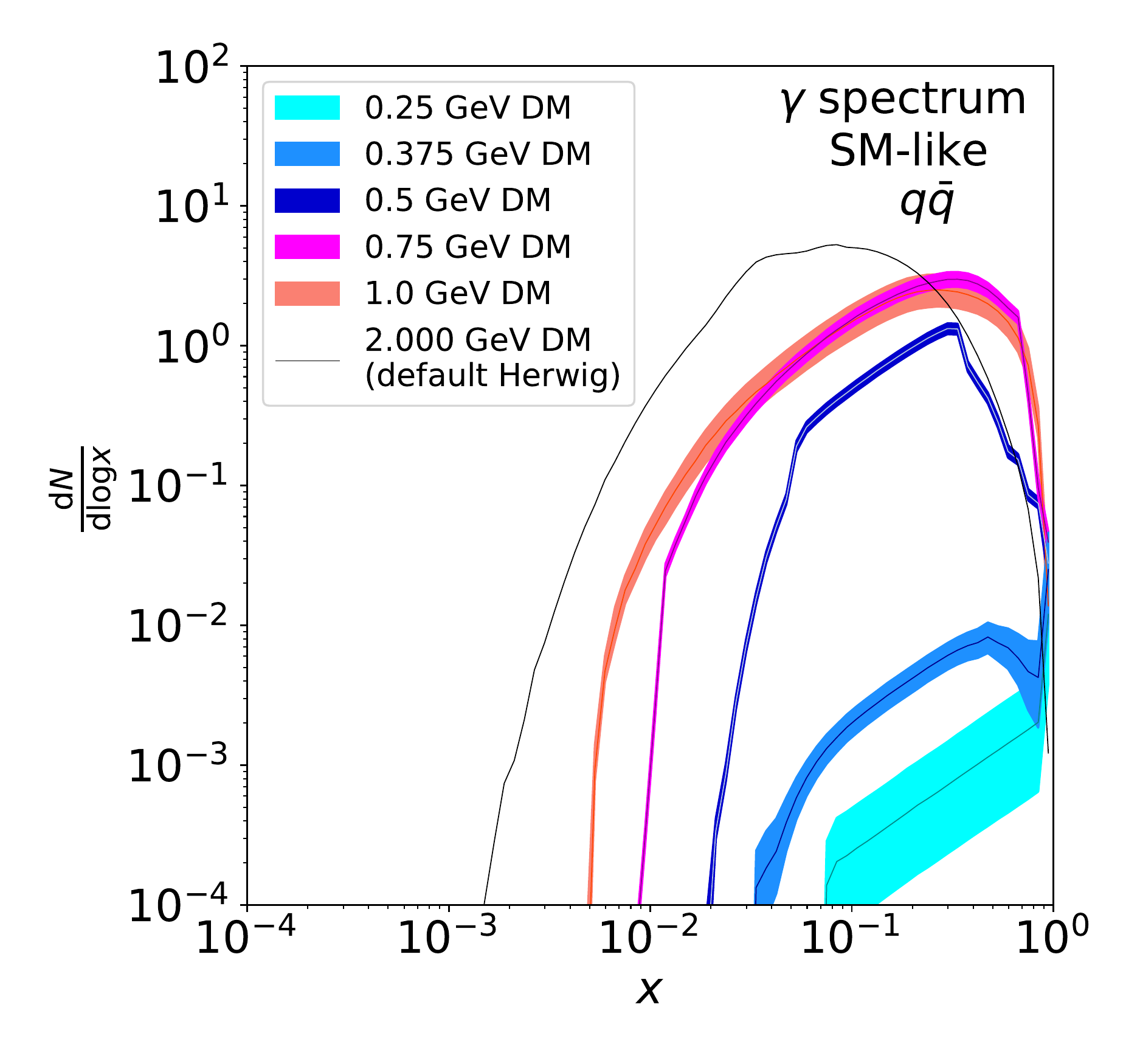}
\includegraphics[width=0.49\textwidth]{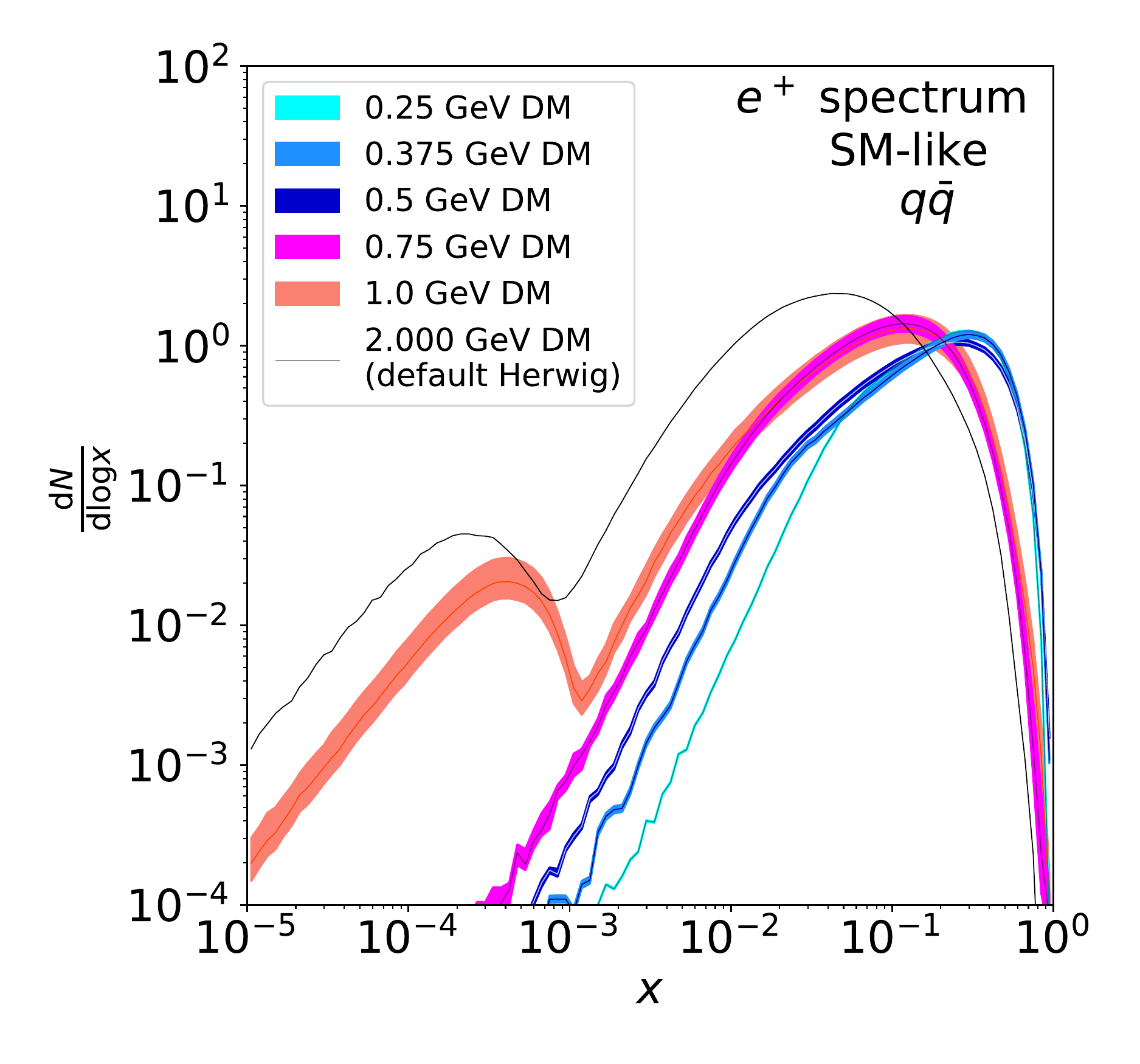}\\
\includegraphics[width=0.49\textwidth]{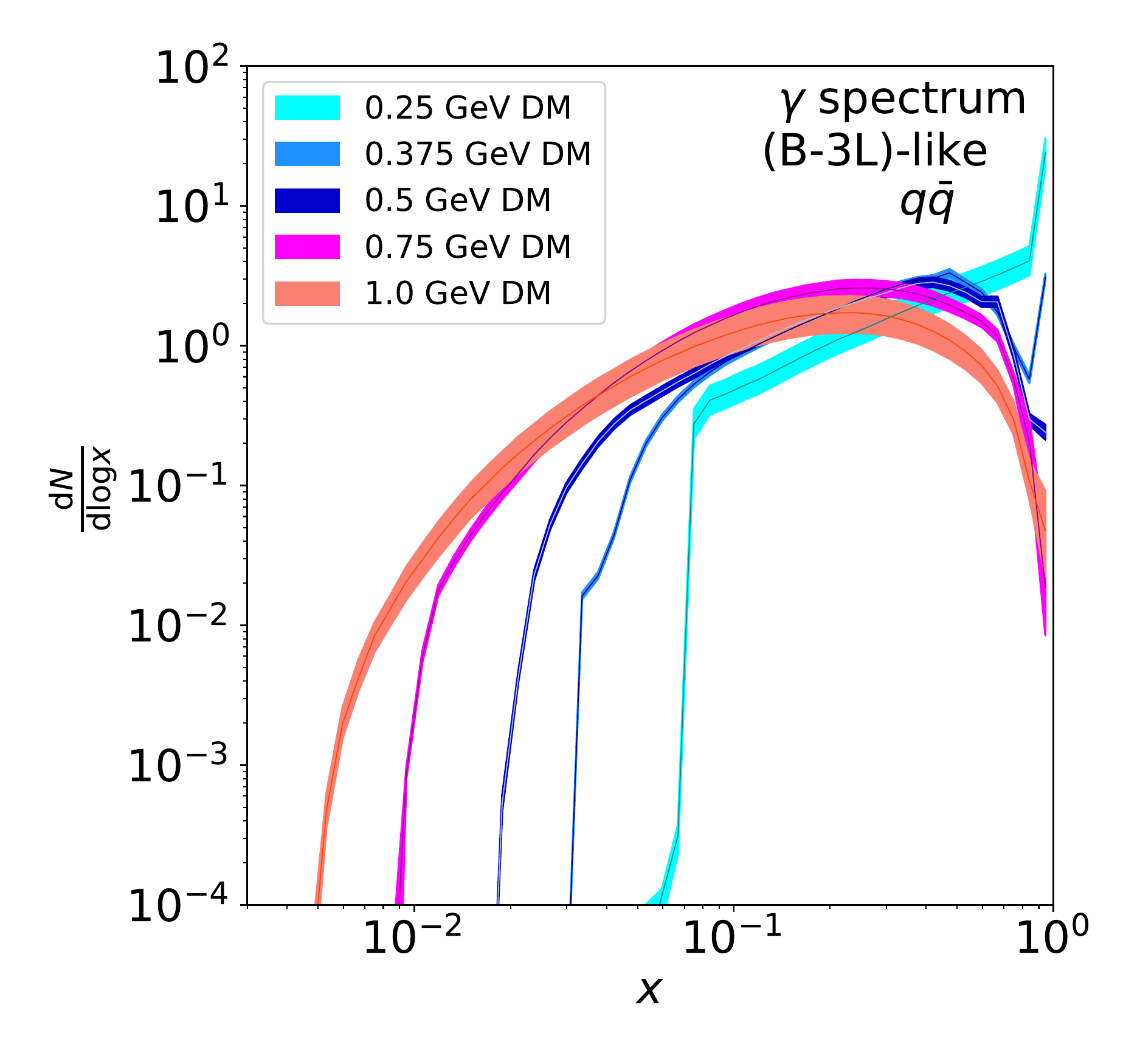} 
\includegraphics[width=0.49\textwidth]{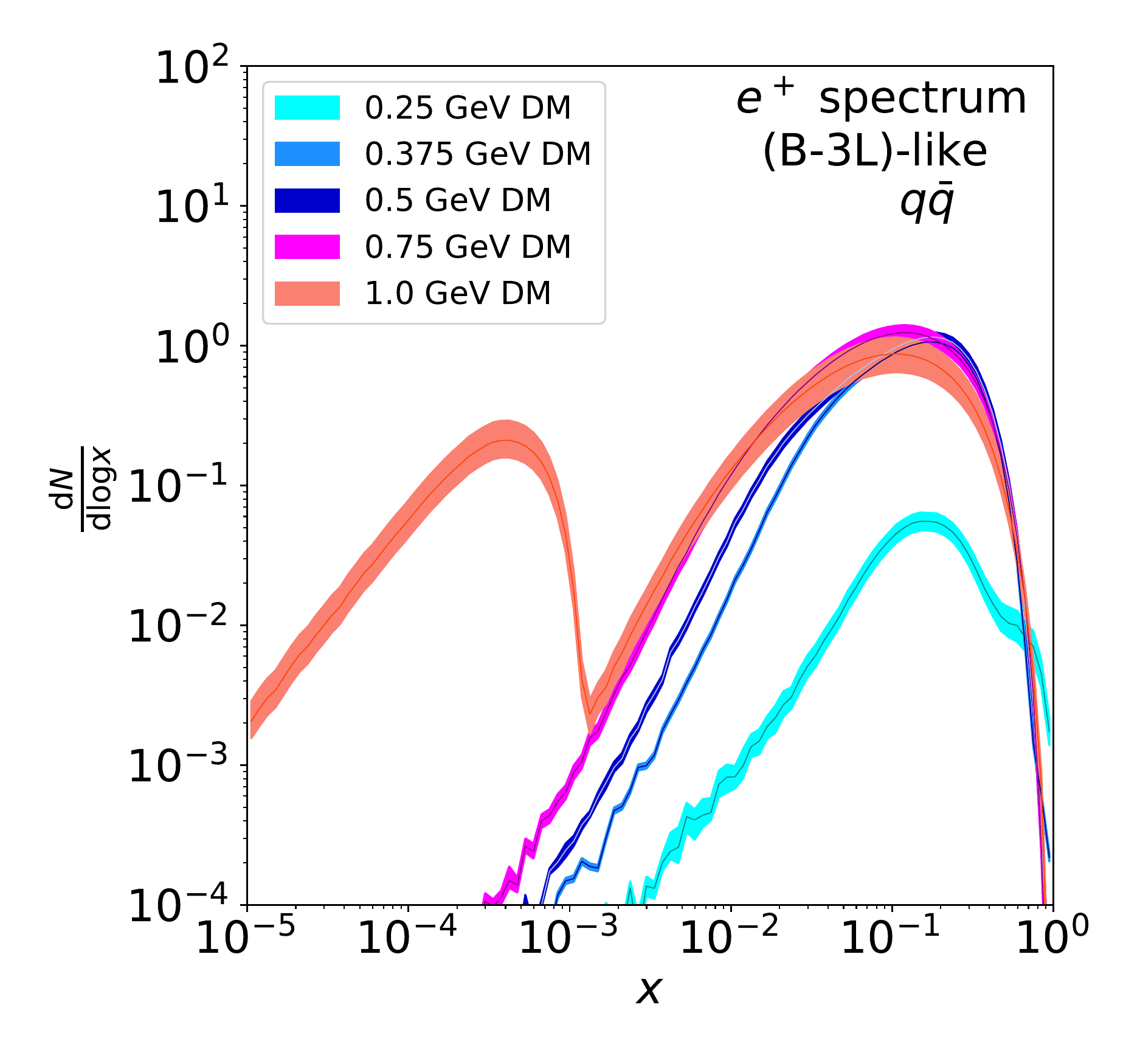}
\caption{Photon and positron spectra $dN /d \log x$ with $x =
  E_\text{kin}/m_{\chi}$ for $m_{\chi}=0.25~...~2$~GeV from $u,d,s,$ quarks
  with SM-like and ($B-3L$)-like couplings with
  uncertainty bands.}
\label{fig:uncertainties_cons}
\end{figure}

Having fixed the hadronic current by low-energy $e^+e^-$ measurements, it is straightforward to modify the left part of the diagram in Fig.~\ref{fig:LEprocess} to fully describe the DM annihilation process. With our \textsc{Herwig}-based implementation\footnote{available at \url{https://herwig.hepforge.org/}}, we can describe every DM model with a fermionic DM candidate and a dark vector mediator.
Hidden sector DM models in the sub-GeV range often include an additional $U(1)$ gauge symmetry~\cite{Foldenauer:2019vgn,Langacker:2009im}. Prominent examples are hidden, or dark, photons and $B-L$ models~\cite{Bauer:2018onh}. Since our study is based on $e^+e^-$-data, we are limited to vector mediator models, and start with an additional $U(1)$ gauge boson $Z'$. For indirect detection searches the annihilation process 
\begin{align}
\chi \chi \to Z' \to \text{SM} 
\end{align}
has a center-of-mass energy of $\sqrt{\hat{s}}=2m_{\chi}$. Hence, depending on the DM mass, different channels will take part in the annihilation depending on the threshold energies listed in Tab.~\ref{tab:channels}. As a model input, we use the DM and vector mediator masses and the couplings of the vector mediator to the DM and SM sector. The coupling strength of the DM to the mediator can be chosen arbitrarily if we are only interested in the form of the energy spectra from hadronic final states. Furthermore, we assume an approximate on-shell annihilation $m_{Z'}\approx 2m_\chi$. Along with the DM mass, the most crucial model-dependent input to the energy spectra is the coupling strength of the mediator to quarks. As seen in Eq.~\eqref{eq:current}, it determines the contribution of $\rho$ meson, $\omega$ meson and $\phi$ meson resonances. We will focus on SM-like couplings $a_{u}=2/3$, $a_{d,s}=-1/3$ as in dark photon models, and $a_q=1/3$ as in any anomaly-free $B-L$ model. Note that in the latter case the $I=1$ current vanishes. In Fig.~\ref{fig:uncertainties_cons}, we show the photon and positron spectra for decreasing DM masses from $m_\chi=2$~GeV to $250$~MeV.

Like in any hadronic process, most photons and positrons come from neutral and charged pion decays. For $\hat{s}\lesssim 1$~GeV, these pions are not only the end product of  hadron decay chains, they are also produced directly. In fact, $2\pi,3\pi, 4\pi$ final states are expected to give leading contributions to hadronic annihilations. Besides, not only pions but also photons can be directly produced, for instance in
\begin{align}
\chi\chi \to \pi^0\gamma, \eta\gamma \; ,
\end{align}
and can contribute to very distinct photon spectra.

Whereas for $m_\chi > 1$~GeV all hadronic states listed in Tab.~\ref{tab:channels} contribute to the round shape of the energy spectrum, for lower energies only photons and positrons from specific processes give a characteristic shape of the spectrum as depicted in Fig.~\ref{fig:uncertainties_cons}. At the two-kaon threshold around $\hat{s}=2m_\chi \sim m_\phi \sim 1$~GeV, we expect that most photons come from the consecutive decays of kaons to pions to photons. This leads to a triangular spectrum. For 250~MeV DM, the only remaining annihilation channels are
\begin{align}
\chi \chi \to \pi^0\gamma, \pi\pi, 3\pi \; ,
\end{align}
where the $\pi^0\gamma$ gives a characteristic sharp peak at around $x=E_{\rm kin}/m_\chi\approx 1$ from the direct photon production. For $B-L$-like couplings with the absence of the $\pi\pi$ channel, the $\pi^0\gamma$ becomes more dominant and lifts the photon spectrum compared to the SM-like case. 

This feature in the photon spectrum is accompanied by a drop in the positron spectrum in the $B-L$ case, when the $\pi\pi$ channel as the main source of positrons drops out.  The basic shape of the positron spectra is the same for almost all energies since all positrons are produced through pion decays rather than directly. The position of the peak of the positron spectrum depends on the stage at which the positrons are produced in the hadron decay chain. For that reason, the spectrum is shifted towards $x=1$ for lower DM masses, where charged pions are produced directly as part of $\pi\pi, 3\pi, 4\pi$ final states and yield positrons in their subsequent decay. The only characteristic peak at around $x\sim 10^{-4}$ arises from the positron production in neutron $\beta$-decays above the neutron pair production threshold.

The main source of uncertainties on the energy spectra are coming from the fits to $e^+e^-$-data. We propagate the uncertainties of the fit parameters through the hadronic currents into the energy spectra. The error of a given spectrum is then dominated by the uncertainty of the leading channel at the corresponding energy. Generally speaking, we obtain smaller uncertainties on the fit parameters for precisely measured channels, like for instance $\pi\pi, 3\pi, 4\pi$. If those channels are dominating the spectrum, the error bands are small. As seen in Fig.~\ref{fig:uncertainties_cons}, this yields large uncertainties in the photon spctrum for the poorly measured dominant $\pi^0\gamma$  channel at $m_\chi=250$~MeV. For the positron spectrum large uncertainties are especially observed in the neutron-$\beta$-decay peak inherited from the poorly measured $n\bar{n}$ channel. Other uncertainties such as parametric uncertainties of the underlying theory have not been taken into account by this first analysis. In~\cite{Plehn:2019jeo}, we show two ways of propagating the fit uncertainties to the spectra and expect that the real error bands including the full error budget are going to be somewhere between those results presented in~\cite{Plehn:2019jeo}. For a more detailed discussion about the uncertainties, we refer to Ref.~\cite{Plehn:2019jeo}.

\section{Moving towards Indirect Detection Constraints on sub-GeV DM}

With the coverage of MeV-range DM annihilations into hadrons, indirect detection constraints on hidden sector DM can be fully explored with \textsc{Herwig} for vector mediator models. For example in Ref.~\cite{Bernreuther:2020koj}, robust CMB constraints are set on dark photons by calculating the effective fraction of injected energy into the CMB per DM annihilation. To study standard indirect detection constraints, \textsc{Herwig}-based energy spectra have been provided for the DarkSUSY DM tool\footnote{available at \url{https://darksusy.hepforge.org/download.html}}. DarkSUSY can interpolate between tabulated energy spectra to photons, positrons, neutrinos and anti-protons for a grid of DM masses. As described above, the energy spectra strongly depend on the way the dark mediator couples to quarks, since some parts of the hadronic current might vanish. By using tabulated energy spectra, we have to restrict ourselves to some exemplary models. Just like above, we choose to provide two of the most studied models in the sub-GeV range, namely the dark photon model with SM-like couplings and $B-L$ models with same couplings to all quarks. The generation of spectra within DarkSUSY serves as a very user-friendly and fast option of obtaining energy spectra in the sub-GeV range. 
The \textsc{Hazma} DM tool is a python package to calculate energy spectra and indirect detection constraints for arbitrary but fixed dark mediator couplings to the DM and quarks. Based on leading-order chiral perturbation theory, a wide range of vector and scalar mediator models can be covered for DM candidates with masses up to 250~MeV. Beyond leading-order, we are currently working on an implementation of form factors to extend the \textsc{Hazma} reach to a few GeV and close the MeV gap for indirect detection DM tools.

\section{Summary}

Indirect detection searches are expected to provide leading constraints on hidden sector DM models with sub-GeV DM masses. Nevertheless, DM annihilations to hadronic final states have been largely unexplored in that energy range. We provide DM annihilation spectra into hadronic final states for vector mediator models in a DM mass range from the 2-pion threshold up to 5~GeV. Our new \textsc{Herwig}-based implementation closes the MeV-gap between standard \textsc{Pythia}-based tools such as \textsc{PPPC4DMID}, \textsc{MicrOMEGAS}, \textsc{MadDM}, or \textsc{DarkSUSY} and the spectra below 250 MeV as provided by \textsc{Hazma}.

We furthermore provide tabulated \textsc{Herwig} energy spectra for \textsc{DarkSUSY} for the most prominent vector mediator models in the sub-GeV range, namely for dark photon and $B-L$ models. Along with the work in progress implementation into \textsc{Hazma}, this opens up new paths to explore indirect detection searches for sub-GeV hidden DM models.

\section*{Acknowledgements}

First of all, the author wants to thank the organizers of the TOOLS for High Energy Physics and Cosmology 2020 conference for the opportunity to present this work. Furthermore, the author is very grateful to Tilman Plehn and Peter Richardson for collaboration on this work and to Oscar Éboli and Tilman Plehn for valuable comments on the manuscript. The author acknowledges financial support from FAPESP under the contract 2020/10004-7 and from the Graduiertenkolleg \textit{Particle physics beyond the
  Standard Model} (GRK 1940).
\bibliographystyle{tep}
\bibliography{literature}

\end{document}